\documentclass[%
 reprint,
 amsmath,amssymb,
 aps,
]{revtex4-2}
\usepackage{tikz}
\usepackage[margin=1in]{geometry}
\usepackage{graphicx}
\usepackage{dcolumn}
\usepackage{bm}
\usepackage{color}

\begin{document}

\preprint{APS/123-QED}

\title{How to Forage for a Mate?\\}

\author{Daniel T. Bernstein}
 \affiliation{Lewis-Sigler Institute, Princeton University\\ 
Center for the Physics of Biological Function, Princeton University }

\author{Ahmed El Hady}
\affiliation{Centre for Advanced Study of Collective Behaviour, University of Konstanz 
 \\
 Department of Collective Behaviour, Max Planck Institute of Animal Behaviour, Konstanz 
}%

\date{\today}

\begin{abstract}
Foraging is a central decision-making behavior performed by all animals, essential to garnishing enough energy for an organism to survive. Similarly, mating is crucial for evolutionary continuity and offspring production. Mate choice is one of the central tenets of sexual selection, driving major evolutionary processes, and can be regarded as a decision-making process between potential mating partners. Often researchers have used coarse-grained models to describe macroscopic phenomenology pertaining to mate choice without detailed quantitative mechanisms of how animals use individual and environmental signals to guide their mating decisions. In this letter, we show that mate choice can be cast as a foraging problem, and we present an analytically tractable optimal foraging-inspired mechanistic theory of decision-making underlying mate choice. We begin from the premise that deciding upon which partner with which to mate is at its core a stochastic decision-making process. Agents adopt a variety of decision strategies, tuned by decision thresholds for leaving or committing to a mate. We find that sensitive leaving thresholds are favored independently of signal availability in the population. By contrast, optimal thresholds for committing to a mate depend upon signal availability in the population, with signal-rich populations generally favoring less eager strategies compared to signal-poor populations.

\end{abstract}
\maketitle
\section{\label{sec:Intro} Introduction}
For sexually reproducing organisms with significant social or biophysical constraints on how many offspring they can have, and on the number of mates with which they can have them, choosing a mate becomes a salient problem \cite{andersson2006sexual}. Any given organism has mating preferences, but determining the presence and abundance of preferred traits in a potential mate takes time––time which is precious to an organism under the pressures of social competition and survival threats \cite{buss2019mate, ryan2013perceptual}.\\
Such a picture of the mate choice problem exhibits strong parallels to the patch foraging problem. In patch foraging models, animals search for food in discrete patches, and have the option to move between patches if the energetic cost of staying in a low-yield patch exceeds the cost of travelling to a new patch \cite{davidson2019foraging, bidari2022stochasticb,kilpatrick2021uncertainty, mcnamara1982optimal}. Agents in these models, depending on environmental statistics, can adopt a variety of decision strategies: for example, minimizing the time it takes to arrive at a high-yielding patch, or maximizing the long term reward rate, spending more time in a high-yielding patch. \\
In this paper, we develop a simple, mechanistic theory of mate choice in the case of an agent choosing between two potential mates. Our theory frames mate choice as a foraging problem: agents accumulate evidence about how well a potential mate aligns with its preferences via discrete signals, and choose to stay with or leave said mate based on the frequency of these signals. Our model differs from traditional foraging theory and similar exploration-exploitation frameworks in that agents must explicitly choose to commit to a single mate \cite{scott2010modern,davidson2019foraging}. In this sense, unlike  traditional foraging behaviors, evidence accumulation and reward-seeking are decoupled, and agents who stay indefinitely with a preferred mate without committing to that mate receive no reward.

As we will show, although both the commitment threshold and leaving threshold of a strategy determine the average reward incurred, the optimality conditions for the leaving threshold are largely independent of signaling rate in the population, unlike conditions for the commitment threshold. This theoretical model can be used to generate hypotheses about mechanistic underpinnings of mate choice across a variety of animal species. 

\begin{figure}
    \centering
    \includegraphics[width=1.0\linewidth]{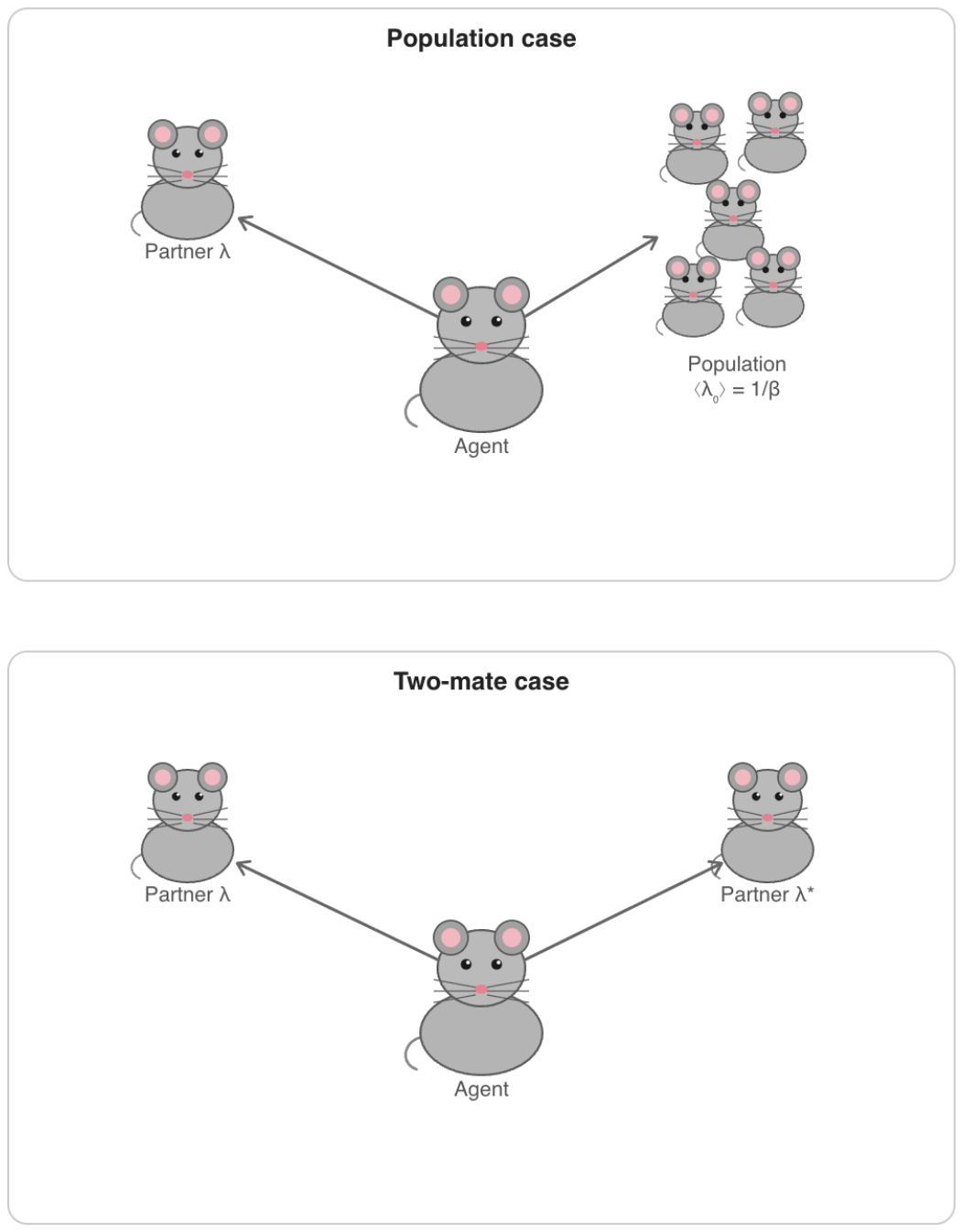}
    \caption{ \textbf {Schematic of mate choice model}. The agent must choose between its current partner and the population (one-mate case, upper panel) or between its current partner and another potential mate (two-mate case, lower panel).}
    \label{fig:placeholder}
\end{figure}

\section{\label{sec:Model} Model Framework}
We consider an agent-based model in which potential mates are characterized by a fitness parameter $\lambda$ indicating alignment to the agent's preferences. This fitness is the mean of a Poisson process: while an agent is spending time with a potential mate of fitness  $\lambda$, it receives Poisson-distributed positive signals from that mate with mean frequency  $\lambda$.



Suppose our agent is currently courting a potential mate of fitness $\lambda$, drawn from a population of fitness distribution $p(\lambda_0)$. In order to compare the current mate to the population, we track the following log-likelihood:
\begin{equation}\label{rho_eq}
\rho(t) = \int_0^{\infty}p(\lambda_0)\log[\frac{P(\lambda>\lambda_0|x(t))}{P(\lambda<\lambda_0|x(t))}]d \lambda_0
\end{equation}

Such a likelihood is typical of Bayesian approaches to foraging and decision-theoretic models \cite{karamched2020bayesian,biernaskie2009bumblebees,trimmer2011decision}

The agent's mating strategy is defined by two thresholds, $\vec{\theta}=(\theta_l$,$\theta_s$), which determine thresholds for leaving and staying with a potential mate, respectively. The leaving threshold determines conditions for which the agent will decide to leave the mate whom they have been courting. The staying threshold determines conditions for which the agent decides to commit to the mate whom they have been courting. More precisely, if $\rho(t)\geq\theta_s$ at any time, the agent commits to the current mate, whereas if $\rho(t)\leq\theta_l$, the agent leaves the current mate. As such, $\theta_\ell$ serves as a sort of measure of a strategy's ``fickleness": the smaller the value of $|\theta_\ell|$, the more readily will the agent leave a mate who is not signalling with sufficient frequency. Likewise, $\theta_s$ measures ``eagerness": low values of $|\theta_s|$ correspond to strategies for which an agent will more readily commit to a mate which is putting out plentiful signals.

Additionally, we consider a signal at which time mating no longer becomes available, whether due to death, competition, et cetera. We choose this signal to be Poisson with scale $\eta$, inspired by the exponentially distributed mortality rate of many prey animals.

The utility of choosing a mate with fitness $\lambda$ is given by the reward function $R(\lambda)$. We start by taking $R(\lambda) = c_2(\frac{e^{c_1 \lambda}-e^{-c_1 \lambda}}{e^{c_1\lambda}+e^{-c_1 \lambda}})$.
\subsection{One-Mate Case}
Consider the case in which the agent is presented with a potential mate and must decide between staying with this mate and leaving to explore the rest of the population. To determine an optimal strategy for a given fitness distribution $P(\lambda)$, we consider the objective function
\begin{equation}\label{single_decision}
   V(\vec{\theta})= \int^{\infty}_0{d\lambda P(\lambda)\left[ P_s(\lambda|\vec{\theta})R(\lambda)+P_l(\lambda|\vec{\theta})R_l(\vec{\theta})\right]}
\end{equation}

where $P_s$ and $P_l$ are staying and leaving probabilities, given the strategy, the fitness value, and the mortality rate, and $R_l$ is a reward assigned to the leaving choice. If we consider a Poisson-distributed (and thus memoryless) mortality rate, then we can impose the self-consistency requirement that $$ R_l(\vec{\theta}) =V(\vec{\theta})$$

Rearranging, we get 

\begin{equation}
    V(\vec{\theta}) = \frac{\int_0^\infty P(\lambda)P_s(\lambda|\vec{\theta})R(\lambda)d\lambda}{1-\int_0^\infty P(\lambda)P_l(\lambda|\vec{\theta})d\lambda}
\end{equation}


\subsection{Two-Mate Case}
Consider the case where the agent must choose between two different mates drawn from the same population, whose true fitnesses are $\lambda$ and $\lambda^\star$. Each time the agent makes a leaving decision, it switches which mate it is observing. When the agent decides to leave a partner, it updates its prior on its fitness according to the accumulated evidence.
We start with the prior $p_0(\lambda_0) = f(\alpha_0=1,\beta_0)$ on the distribution of fitness in the population, where $f(\alpha,\beta)$ is a gamma distribution with shape $\alpha$ and rate $\beta$: in other words, we start with an exponential prior. So, if the agent receives $n$ positive signals over an interval of time $T$ before deciding to leave a mate whose original prior was $p_{old} = f(\alpha,\beta)$, then the update rule for the prior is: $$p_{new}(\lambda) = \frac{p_{old}(\lambda)p_{old}(x(t)|\lambda)}{p(x(t))}=f(\alpha+n,\beta+T)$$

What becomes of the form of \eqref{rho_eq}? Putting 
$K(t) \equiv \int_{0}^{t}{x(t')dt'}$, $$P(\lambda>\lambda_0|x(t)) = \frac{1}{p(x(t))}\int_{\lambda_0}^{\infty}{p(\lambda)\lambda^{K(t)}e^{-\lambda t}d\lambda}$$

So:
$$P(\lambda>\lambda_0|x(t)) = \frac{1}{p(x(t))}\int_{\lambda_0}^{\infty}{[\frac{\lambda^{\alpha-1}e^{-\beta \lambda}\beta^{\alpha}}{\Gamma(\alpha)}]\lambda^{K(t)}e^{-\lambda t}d\lambda}$$

$$P(\lambda>\lambda_0|x(t)) = \frac{\beta^\alpha \int_{\lambda_0(\beta + t)}^{\infty}{e^{-\nu}\nu^{\alpha+K(t)-1}d\nu}}{p(x(t))\Gamma(\alpha)(\beta+t)^{\alpha+K(t)}}$$

\begin{equation}
\begin{split}
&  P(\lambda>\lambda_0|x(t)) = \\
&\quad \frac{\beta^\alpha}{p(x(t))\Gamma(\alpha)(\beta+t)^{\alpha+K(t)}}\Gamma(\alpha+K(t),\lambda_0(\beta+t))
\end{split}
\end{equation}

Where $\Gamma(,)$ is the upper incomplete gamma function. Following a similar process for $P(\lambda<\lambda_0|x(t))$ but arriving at the lower incomplete gamma function $\gamma(,)$, we get
\begin{equation}
\rho(t) = \int^{\infty}_0\log[\frac{\Gamma(\alpha+K(t),\lambda_0(\beta+t))}{\gamma(\alpha+K(t),\lambda_0(\beta+t))}]p(\lambda_0|\beta^\star,\alpha^\star)d\lambda_0    
\end{equation}

\section{\label{sec:Model} Results}
\subsection{One-Mate Case}
\begin{figure}
    \centering
    \includegraphics[width=\linewidth]{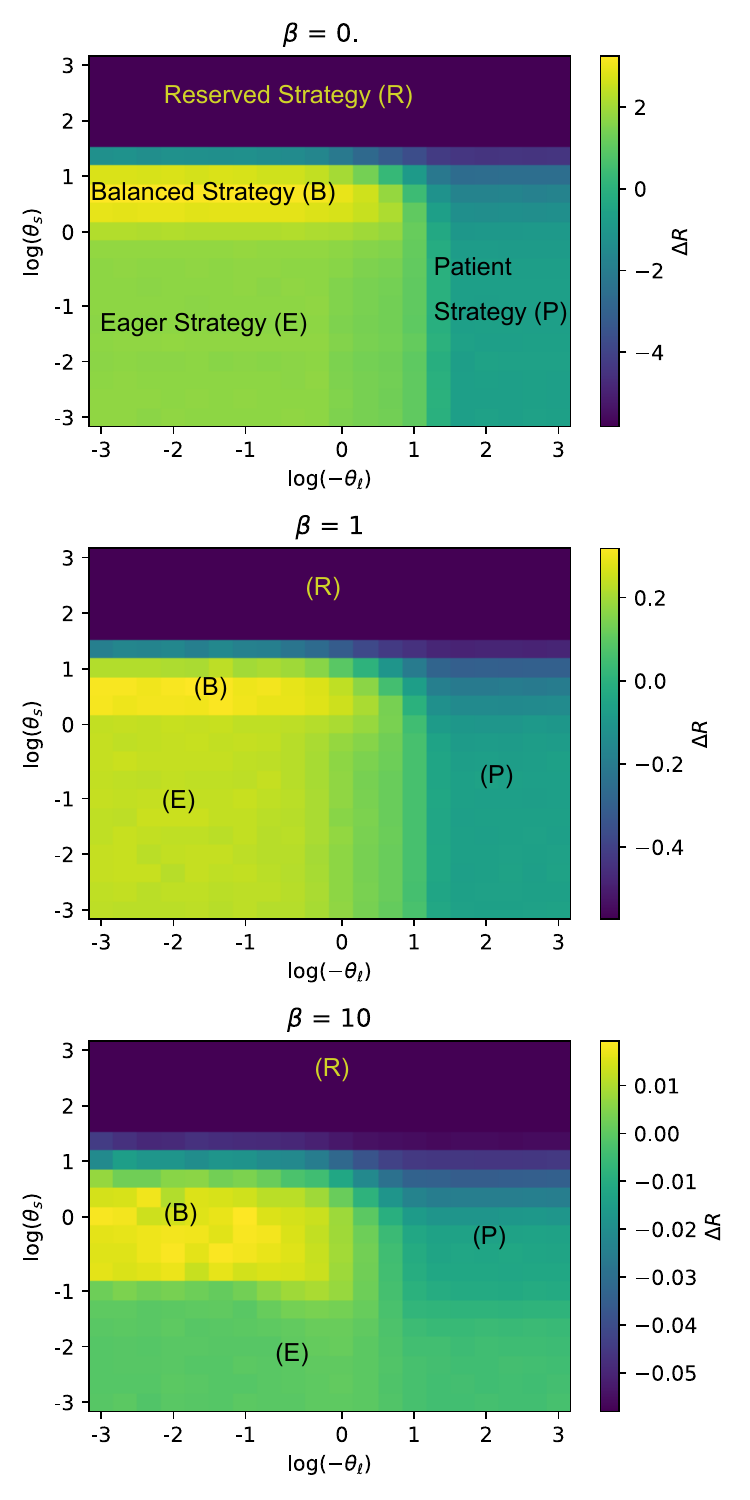}
    \caption{ \textbf {Emergence of different strategies in the one - mate case}, with differing average reward relative to population baseline ($\Delta R(\vec{\theta}) = V(\vec{\theta})-\langle R(\lambda)\rangle$, with $R(\lambda)= \frac{1}{\beta}\tanh(\beta\lambda)$).  There is an optimal balanced strategy which performs distinctly better than the increasingly suboptimal eager, patient, and reserved strategies, which respectively correspond to excessively low staying thresholds, excessively high leaving thresholds, and excessively high staying thresholds. As information becomes less available ($\beta \sim 10$), the eager and patient strategies become equivalently suboptimal, and the balanced strategy becomes the only strategy to perform above random chance. Behavioral strategies are abbreviated as follows: Reserved (R), Balanced (B), Eager (E) and Patient (B) strategies }
    \label{fig:beta_heatmap_one}
\end{figure}

Figure \ref{fig:beta_heatmap_one} details the average reward yielded by different strategies in the one-mate case in three different regimes of information rate. We assume an exponential prior (rate of $\beta$) as in the two-mate case, and we choose $R(\lambda) = \frac{1}{\beta} \tanh(\beta \lambda)$. We fix the mortality rate to be $\eta = 100$. 
When the potential mates are signaling at a high rate relative to $\eta$ ($\beta = 0.1$), the most optimal strategies are those for which $|\theta_\ell|$ is low ($\log (-\theta_\ell)\lesssim 0 $), whereas $\log (\theta_s)\gtrsim  0 $. We observe four phases within which average reward remains relatively constant: the aforementioned optimal region, a suboptimal ``eager" region of lower $\theta_s$ values, a further suboptimal ``patient" region of higher $|\theta_\ell|$ values, and a dismal ``reserved" region of excessively high $\theta_s$ values.

In the ``medium-information" case ($\beta = 1$), the sharp boundary softens between the optimal and eager regions. In the low-information regime ($\beta = 10$), we see the optimal region grow and shift downwards, and the phase boundary separating the eager and patient regions shifts from a vertical to a horizontal one.

Interestingly, for a given leaving threshold $\theta_\ell$, the relationship between information availability ($\beta$) and the average baseline-adjusted reward $\Delta R$ follows a power law: see Figure \ref{fig:powerlaw}. For small $\theta_\ell$, the power-law exponent is approximately $-1$ (see Appendix A for an analytical justification of this result). This suggests that, for small leaving thresholds and reasonable staying thresholds, we expect the average increase in reward over baseline for a given strategy to scale linearly with the average fitness of the population.

\subsection{Two-Mate Case}
\begin{figure}
    \centering
    \includegraphics[width=\linewidth]{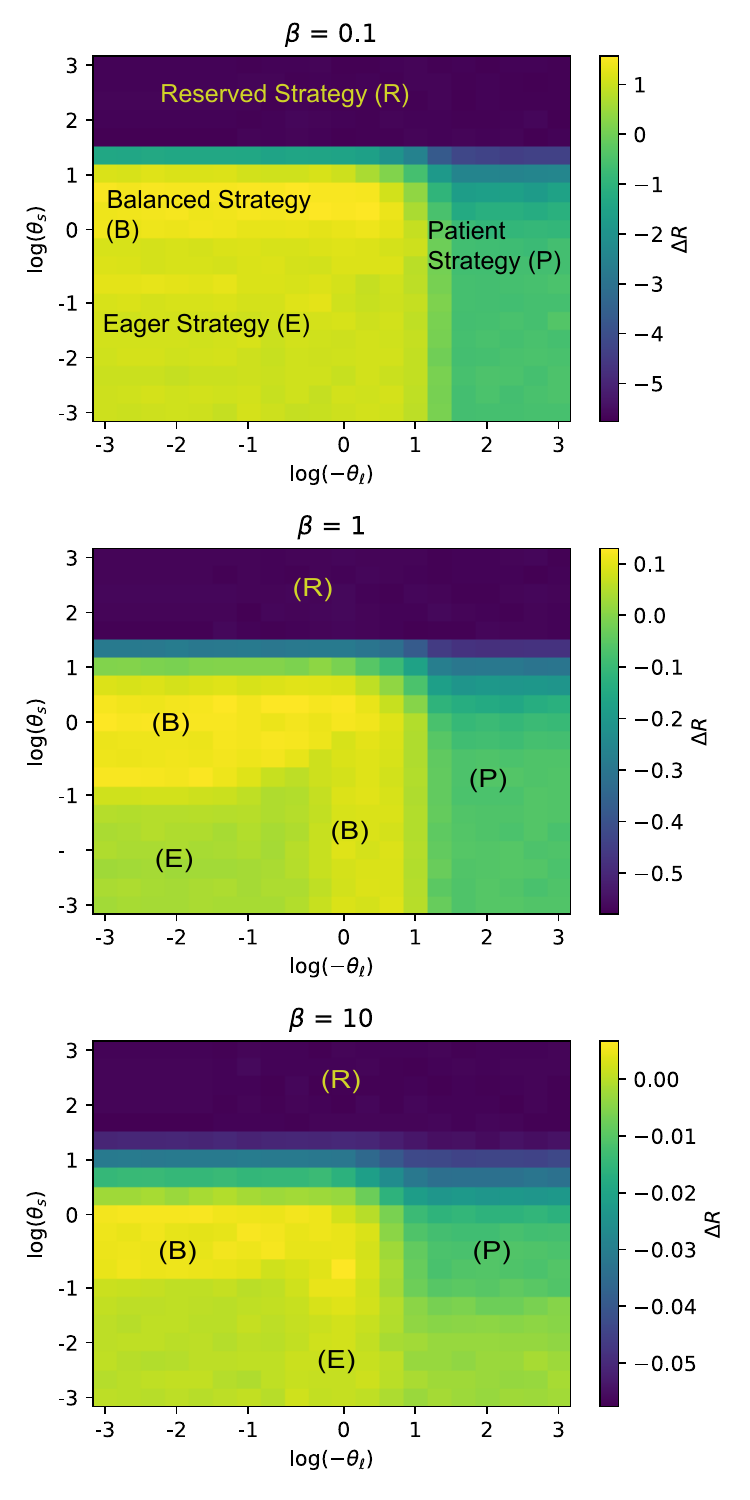}
    \caption{\textbf {Emergence of different strategies in the two - mate case}. Strategies change with information rate in a similar manner to the one-mate case, with the notable exception that the region of optimal balanced strategy shifts in the intermediate information ($\beta$ = 1) case. Behavioral strategies are abbreviated as follows: Reserved (R), Balanced (B), Eager (E) and Patient (B) strategies}
    \label{fig:two_heat}
\end{figure}
The results of the two-mate case suggest a similar relationship between the reward rate and the phase diagram of strategies as in the one-mate case. Once again, the optimal region of strategies for high reward rate is characterized by $ 0 \lesssim \log\theta_s\lesssim 1$ and $\log -\theta_\ell \lesssim 1$. This region expands around the eager region in the medium-information regime. As in the one-mate case, the optimal region shifts downward in the low-information regime, and the boundary shift between the optimal, eager, and patient regions in the low-information regime resembles the shift we observe in the one-mate case (although here the eager region is less suboptimal). 
\begin{figure}
    \centering
    \includegraphics[width=\linewidth]{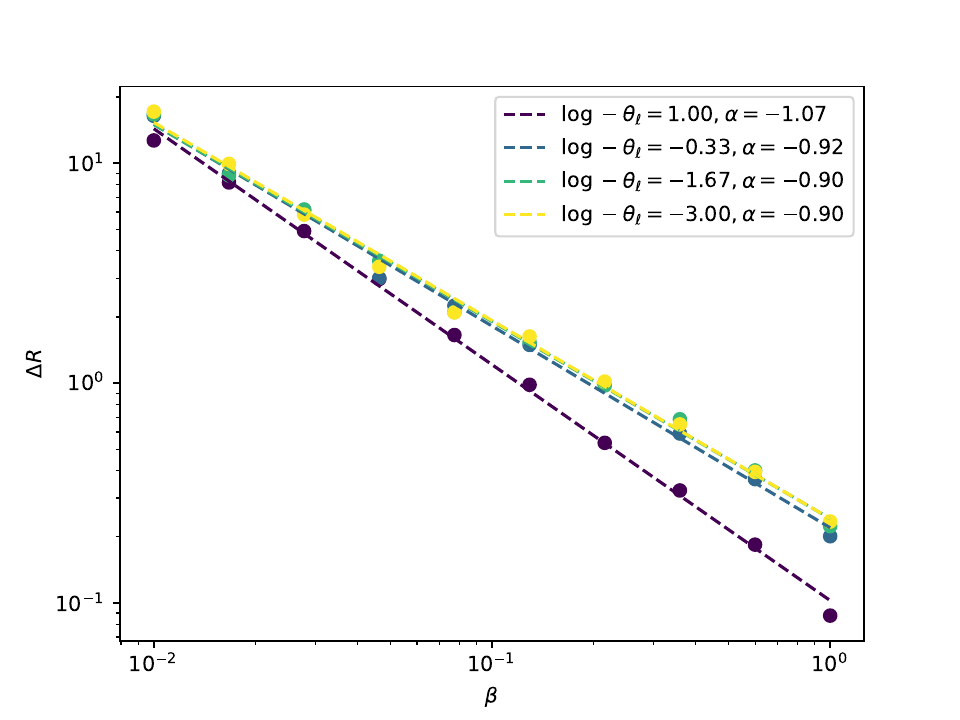}
    \caption{ \textbf{Emergent scaling relationship}. Power laws between $\Delta R$ and $\beta$, set by $\theta_\ell$. For small values of $\log(-\theta_\ell)$, we observe a power-law exponent $\alpha \approx-0.9$, consistent with our analytical estimation of a $\frac{1}{\beta}$ scaling.}
    \label{fig:powerlaw}
\end{figure}
\section{\label{sec:Conclusions} Conclusions}
We have introduced a simple model of mate choice, inspired by optimal foraging theory, for monogamous organisms which must choose a partner in information-limited environments. Unlike the case of patch foraging, in which optimal strategies are specified by the conditions under which organisms decide to leave a patch, our model considers both the leaving decision and a commitment decision with regard to a given potential mate. Although there has been previous qualitative proposals casting mate choice as a potential foraging problem \cite{louapre2015male}, our study is the first of its kind to quantitatively formalize it.  

We find that, in the case of populations where fitness is exponentially distributed, and the mortality rate of organisms is also assumed to be exponential, organisms on average are rewarded for ``fickle" strategies (strategies with a sensitive leaving threshold), and the optimal threshold for the staying decision depends upon the average signaling frequency in the population. Furthermore, we find that the leaving threshold sets a power-law relationship between average signaling frequency (mate quality) in the population and the average expected reward, with more sensitive thresholds corresponding to more favorable exponents.

Across many animal species, there are increasing evidence that individuals integrate sensory cues and other signaling to guide their mate choice strategies: for example, Drosophila courtship involves integrating acoustic communication signals \cite{day2026neural, pang2025inferring} , Canaries mating involves integrating environmental cues and songs produced by potential mates \cite{trosch2017genes} and Marmoset mating behaviors involve the integration of stereotypical mating behaviors such as grooming and approach \cite{drazan2025improving}. Interestingly, human online dating behavior can also be regarded as foraging involving differential pools of potential partners where the choice dynamics are modulated by multiple factors from mate characteristics to the nature of the online setting \cite{jung2022secret}.

We note that, although our model does not capture many important aspects of mate choice across populations, as we have collapsed the multitude of signals an agent receives into an abstract quantity, it lends itself naturally to a variety of extensions. Among these extensions are the use of different population fitness distributions, the incorporation of cognitive mechanisms underlying mate choice, \cite{miller1998mate, castellano2010computational} multiple sensory cues used in mate choice \cite{candolin2003use}, perceptual biases  \cite{ryan2013perceptual}, negative signals of mate quality, deceptive signaling \cite{mccoy2020embryo}, explicit competition between mates \cite{fawcett2003mate}, bidirectional mate choice, statistics of mate spatial movement and the extension of our one-mate and two-mate cases to a general n-mate case. By systematically examining different extensions, we can better understand what socio-ecological factors shape optimal mate choice strategies within populations, and how these strategies become evolutionary stable across organisms which exhibit different social behaviors and different biophysical constraints on mate choice, therefore opening up the space for a quantitative mechanistic framework for sexual selection, one of the foundational pillars of evolutionary theory.

\begin{acknowledgments}
We would like to acknowledge William Bialek, Joshua Shaevitz, Ben Sorkin and all members of the Center for Physics of Biological Function at Princeton University. We would like to also acknowledge Lisa Blum Moyse and all members of the Integrative Biophysics Group at University of Konstanz for helpful discussions and suggestions.
AEH is supported by by the DFG German Research Foundation (EXC 2117-422037984). and Human Frontiers Science Foundation Grant (RGP006/2025). We would like to acknowledge the Konstanz School of Collective Behavior where the idea behind this project was concieved. 
\end{acknowledgments}

\appendix

\section{Power Law}

To understand why a power law arises between $\Delta R$ and $\beta$ when  $\theta_\ell$ is small, we investigate the behavior of $\rho(t)$ for $K = 0$ and $\alpha=\alpha^\star = 1$, $\beta = \beta^\star$, i.e. how $\rho$ evolves in the two-mate case before the first signal arrives. 
\begin{align}
    \rho(t) &=  \int_0^{\infty}{\beta e^{-\beta \lambda_0}\ln(\frac{e^{\lambda_0(\beta+t)}}{1-e^{\lambda_0(\beta+t)}})d\lambda_0}  \nonumber\\
    &= \int_0^{\infty}{\beta e^{-\beta \lambda_0}\ln(e^{\lambda_0(\beta+t)})d\lambda_0} \nonumber\\
    &\phantom{AAA} -\int_0^{\infty}{\beta e^{-\beta \lambda_0}\ln(1-e^{\lambda_0(\beta+t)})d\lambda_0} \nonumber\\
    &= -\frac{\beta + t}{\beta} +\psi(\frac{\beta}{\beta+t}+1)+\gamma \nonumber\\
    &= \psi(\frac{\beta}{\beta+t})+\gamma = \theta_\ell
\end{align}
where $\psi$ is the digamma function and $\gamma$ is the Euler-Mascheroni constant. $\psi$ has an inverse on the interval $(0,1]$, which we denote $\psi^{-1}$, and so we can rearrange and write 
\begin{equation}
   t= \frac{\beta(1-\psi^{-1}(\theta_l-\gamma))}{\psi^{-1}(\theta_l-\gamma)}
\end{equation}

which shows that, for a fixed $\theta_l$, the relationship between the leaving time (given no signal) and the population value of $\beta$ is linear, with coefficient given by $C_\ell=\frac{(1-\psi^{-1}(\theta_\ell-\gamma))}{\psi^{-1}(\theta_\ell-\gamma)}$. For sufficiently small $\theta_l$ (relative to a fixed range of $\beta$), we expect leaving decisions to occur before the first signal is received, so we expect an approximately linear relationship between the value of $\beta$ and the time spent with a mate which the agent will ultimately leave. Likewise, if we set $\theta_s$ small, then if an event occurs before $t$, then it follows that the agent will commit to that mate--but as long as $\theta_s$ is sufficiently small, the exact value of $\theta_s$ is irrelevant.  In this setting, $P(\ell) = (e^{-{\lambda \beta C_\ell}})e^{-\frac{1}{\eta}\beta C_\ell}$ and $P(s) = (1-e^{-{\lambda \beta C_\ell}})e^{-\frac{1}{\eta}\beta C_\ell}$. 

Plugging into 
\begin{align}
\Delta R &= V(\vec{\theta})-\int^{\infty}_0{\beta e^{-\beta \lambda}R(\lambda)d\lambda} \nonumber \\ 
&= \frac{\int_0^\infty P(\lambda)P_s(\lambda|\vec{\theta})R(\lambda)d\lambda}{1-\int_0^\infty P(\lambda)P_l(\lambda|\vec{\theta})d\lambda} -\int^{\infty}_0{\beta e^{-\beta \lambda}R(\lambda)d\lambda} \nonumber \\
&= \frac{\int_0^\infty e^{-\beta \lambda}(1-e^{-{\lambda \beta C_\ell}})e^{-\frac{1}{\eta}\beta C_\ell}\tanh(\beta \lambda)d\lambda}{1-\int_0^\infty \beta e^{-\beta \lambda}(e^{-{\lambda \beta C_\ell}})e^{-\frac{1}{\eta}\beta C_\ell}d\lambda} \nonumber \\
&- \int^{\infty}_0{e^{-\beta \lambda}\tanh(\beta \lambda)d\lambda} \nonumber \\
&= \frac{1}{\beta}(A(\eta,\theta_\ell)),
\end{align}

where 

\begin{align}
&A(\eta,\theta_\ell) = \nonumber\\
&=\left(1 - \frac{\pi}{2} + \frac{C_\ell(\pi - 2) + \pi + (1+C_\ell)(\,\psi\!\left(\frac{1+C_\ell}{4}\right) - \,\psi\!\left(\frac{3+C_\ell}{4}\right))}{2(-1 + (1+C_\ell)\,e^{\frac{\beta C_\ell}{\eta}})}\right)\nonumber
\end{align}
We find that
\begin{equation}
    \ln (\Delta R) = - \ln (\beta) - \ln (A(\eta,\theta_\ell))
\end{equation}

This result agrees with Figure \ref{fig:powerlaw}, which predicts a power law of exponent $\sim-1$ for small $\theta_\ell$.

\bibliography{mateforage}

\end{document}